\documentclass[twoside,english]{article}
\usepackage[T1]{fontenc}
\usepackage[latin1]{inputenc}
\usepackage{geometry}
\usepackage{amsmath}
\usepackage{amssymb}

\makeatletter
\usepackage{amsfonts}
\usepackage{amssymb}

\usepackage{babel}
\makeatother
\begin{document}

\newcommand{\set}[2]{\left\{  #1\, |\, #2\right\}  }

\newcommand{\cyc}[1]{\mathbb{Q}\left[ \zeta_{#1 }\right] }

\newcommand{\Mod}[3]{#1\equiv#2\, \left(\mathrm{mod}\, \, #3\right)}

\newcommand{\FA}[1]{N}

\newcommand{\FB}[1]{\mathbb{P}\mathcal{K}}

\newcommand{\FC}[1]{\mathcal{G}_{#1}}

\newcommand{\FD}[1]{\Gamma\left(K,\mathfrak{#1}\right)}

\newcommand{\FE}[2]{\ker\pi}

\newcommand{\FF}[2]{}

\author{P. Bantay\\
{\normalsize Inst. for Theor. Phys., Eotvos Univ. Budapest}}

\title{Galois currents and the projective kernel in Rational Conformal Field
Theory}

\maketitle
\begin{abstract}
The notion of Galois currents in Rational Conformal Field Theory is
introduced and illustrated on simple examples. This leads to a natural
partition of all theories into two classes, depending on the existence
of a non-trivial Galois current. As an application, the projective
kernel of a RCFT, i.e. the set of all modular transformations represented
by scalar multiples of the identity, is described in terms of a small
set of easily computable invariants. 
\end{abstract}

\section{Introduction}

A most important characteristic of a (Rational) Conformal Field Theory
is the associated modular representation \cite{Cardy}. Not only does
it prescribe the transformation of the genus one characters of the
primary fields under the modular group $\Gamma\left(1\right)=SL\left(2,\mathbb{Z}\right)$,
but it does also determine the central charge (modulo 8), the fractional
part of the conformal weights, the fusion rules via Verlinde's formula,
etc.\cite{Ver,MS}. All these properties are linked to the fact that
the modular representation of an RCFT provides part of the defining
data of a modular tensor category \cite{Turaev,MTC,FRS}. 

A fundamental property of the modular representation, conjectured
by many authors over the years \cite{Moore,Eholzer,Eh-Sko,BCIR,CG2}
and finally proved in \cite{CMP}, is the congruence subgroup property:
there exists a natural number $\FA{}$ (which turns out to equal the
order of the matrix $T$ representing the Dehn-twist $\tau\mapsto\tau+1$),
such that the kernel of the modular representation contains the principal
congruence subgroup \[
\Gamma\left(\FA{}\right)=\set{\left(\begin{array}{cc}
a & b\\
c & d\end{array}\right)\in\Gamma\left(1\right)}{\Mod{a,d}{1}{\FA{}},\,\,\Mod{b,c}{0}{\FA{}}}\,.\]
In other words all modular transformations which belong to $\Gamma\left(\FA{}\right)$
are represented by the identity. This result has far reaching implications
both in theory and practice, e.g. it is the basis of powerful algorithms
to compute arbitrary modular matrices. 

Once we know that the kernel is a congruence subgroup, we can look
for a complete description of it in terms of a small set of easy-to-compute
invariants. As it turns out, this won't work for the kernel, but the
reason is that the kernel is not the right notion to look at. It is
the projective kernel, i.e. the set of modular transformations represented
by scalar matrices, that is the natural object, and for it one does
indeed get a simple description. Along the way emerges the notion
of Galois currents, which are simple currents \cite{sc1,sc2,sc3}
related to the Galois action \cite{BG,CG1}. The basic result about
Galois currents is uniqueness: a model can have at most one non-trivial
Galois current. This leads at once to the partition of all RCFTs into
two classes, according to whether they have a non-trivial Galois current
or not. As we shall see, while the latter are the generic ones, many
important examples of RCFTs - like the Ising model - fall into the
first class. 

This paper aims to give a brief survey of the above results, concentrating
on the conceptual issues, and leaving the technical details to a separate
publication. Besides introducing the relevant notions and stating
the main results, a couple of simple examples are included in order
to illustrate the general theory.

\section{Modular matrices and the Galois action}

As indicated in the introduction, to each rational CFT is associated
a finite dimensional representation $D$ of the modular group $\Gamma\left(1\right)$.
This representation is most conveniently described by a pair of matrices
$S$ and $T$, which represent the generators $\left(\begin{array}{cc}
0 & -1\\
1 & 0\end{array}\right)$ and $\left(\begin{array}{cc}
1 & 1\\
0 & 1\end{array}\right)$ of the modular group. These matrices have to satisfy the defining
relations \begin{eqnarray*}
STS & = & T^{-1}ST^{-1}\,,\\
S^{4} & = & 1\end{eqnarray*}
of the modular group. It should be noted that the matrices $S$ and
$T$ carry much more information than a mere linear representation
of $\Gamma\left(1\right)$, for the individual matrix elements have
an invariant meaning in themselves, in other words the representation
comes equipped with a distinguished basis formed by the genus 1 characters
of the primary fields. The properties of the matrix elements with
respect to this distinguished basis are summarized in Verlinde's theorem
\cite{Ver,MS}:

\begin{enumerate}
\item The matrix $T$ is diagonal and of finite order;
\item The matrix $S$ is symmetric;
\item The fusion rules of the primaries are given by Verlinde's formula
\[
N_{pqr}=\sum_{s}\frac{S_{p}^{s}S_{q}^{s}S_{r}^{s}}{S_{0}^{s}}\,,\]
where $0$ labels the vacuum (the conformal block of the identity).
\end{enumerate}
One may show that the modular representation is quite special in many
respects, the basic result being the congruence subgroup property
mentioned in the introduction, i.e. that the kernel $\ker D$ of the
modular representation is a congruence subgroup of level $\FA{}$
\cite{CMP}. One may also show that the \emph{conductor} $\FA{}$,
which is equal to the order of the matrix $T$, is bounded by a universal
function of the dimension (= number of primary fields). Moreover,
if we denote by $K$ the projective order of $T$, i.e. the smallest
positive integer such that $T^{K}$ is a scalar matrix, then the integer
$e=\frac{\FA{}}{K}$ is a divisor of $12$ \cite{CMP}. The integers
$e$ and $K$ are important characteristics of the model.

Because the individual matrix elements have an invariant meaning,
it is meaningful to look at their arithmetic properties, which is
the content of the theory of the Galois action \cite{BG,CG1}. The
basic idea is to look at the field $F$ obtained by adjoining to the
rationals the matrix elements of all modular matrices (it is enough
to adjoin those of the generators $S$ and $T$). It turns out that
the field $F$ equals the cyclotomic field $\cyc{\FA{}}$, where $\FA{}$
is the conductor of the model. By a well known result of algebraic
number theory, the Galois group of the cyclotomic field $\cyc{\FA{}}$
over $\mathbb{Q}$ is isomorphic to the group $\FC{\FA{}}=\left(\mathbb{Z}/NZ\right)^{*}$
of prime residues modulo $\FA{}$, where to the prime residue $l\in\FC{\FA{}}$
corresponds the map $\sigma_{l}$ sending $\zeta_{\FA{}}$ to its
$l$-th power $\zeta_{\FA{}}^{l}$. As $\sigma_{l}$ maps $F$ into
itself, it is meaningful to consider its action on the matrix elements
of $S$ and $T$. The basic result \cite{BG,CG1}, which follows from
Verlinde's theorem, is \[
\sigma_{l}\left(S_{q}^{p}\right)=\varepsilon_{l}\left(q\right)S_{\pi_{l}q}^{p}\,,\]
where $\pi_{l}$ is a permutation of the primaries - the \emph{Galois
permutation} associated to $l\in\FC{\FA{}}$ -, while $\varepsilon_{l}\left(q\right)=\pm1$
is a sign. The above Galois permutations define a permutation action
of $\FC{\FA{}}$ on the set of primaries, because they satisfy $\pi_{lm}=\pi_{l}\pi_{m}$.

\section{Galois currents}

According to the traditional definition \cite{sc1,sc2,sc3}, a simple
current is a primary field $\alpha$ whose quantum dimension is $1$,
i.e. such that $S_{0}^{\alpha}=S_{0}^{0}$. From this definition follows
that the fusion product of a simple current $\alpha$ with a primary
$p$ is again a primary field, denoted $\alpha p$, i.e. simple currents
induce permutations of the primaries. Of course, the permutations
arising this way are quite special, and it is a challenging problem
to characterize those permutation groups that may correspond to the
action of simple currents on the primary fields of some RCFT. In any
case, this induced permutation action allows us to consider for each
simple current $\alpha$ the set \[
L_{\alpha}=\set{l\in\FC{\FA{}}}{\alpha p=\pi_{l}p}\,,\]
i.e. the set of prime residues mod $\FA{}$ such that the Galois permutation
$\pi_{l}$ is the same as the permutation induced by $\alpha$ (this
set may be empty). We call a simple current $\alpha$ a \emph{Galois
current} when $L_{\alpha}\neq\emptyset$. The set of Galois currents
may be shown to be closed under the fusion product, in other words
they form a group.

Note that the vacuum is always a Galois current, as it is a simple
current for which the corresponding permutation is the identity, consequently
$1\in L_{0}$. The interesting question is whether there exists models
with non-trivial simple currents, so let's give a couple of simple
examples.

Our first example is the Ising model, the Virasoro minimal model $\mathcal{M}\left(4,3\right)$
of central charge $c=\frac{1}{2}$. As it is well known, this model
has 3 primaries: the vacuum $0$, the energy operator $\varepsilon$,
and the spin operator $\sigma$. One has $K=16$ and $e=3$ in this
model. A simple computation reveals that $L_{\varepsilon}=\left\{ 5,11,13,19,29,35,37,43\right\} $,
i.e. $\varepsilon$ is a non-trivial Galois current. More generally,
one may show that the only Virasoro minimal models with a non-trivial
Galois current are those of the form $\mathcal{M}\left(4,p\right)$
with $p$ odd - the Ising model corresponding to $p=3$ -, in which
case the primary field with Kac-label $\left(3,1\right)$ is the non-trivial
Galois current (for $p>3$).

There are many more examples of theories with non-trivial Galois currents,
e.g. some of the Ashkin-Teller models (= $\mathbb{Z}_{2}$ orbifolds
of the compactified boson, the Galois current being the marginal perturbation),
many WZNW models, products and orbifolds of the above, etc. As these
examples show, Galois currents appear in many types of RCFTs. Closer
examination of the above examples leads to the following observations: 

\begin{itemize}
\item the group of Galois currents is small (in all of the above examples
its order is either 1 or 2);
\item theories with non-trivial Galois currents are sparse: in two-parameter
families of models they form one parameter subfamilies.
\end{itemize}
As it turns out, the above observations reflect general properties
of RCFT, for one may show - the proof is quite lengthy and technical
- that 

\begin{enumerate}
\item If there is a non-trivial Galois current, then it is unique, i.e.
the group of Galois currents has either 1 or 2 elements;
\item If a theory has a non-trivial Galois current, then $e$ is odd and
$K$ is a multiple of $16$.
\end{enumerate}
With respect to this last result, we note that the reverse implication
is by no means true, for there exist examples of RCFTs with odd $e$
and $K$ a multiple of $16$, but no non-trivial Galois current. On
the other hand, in those classes of RCFTs for which one has an explicit
expression for $e$ and $K$, those with odd $e$ and $K$ a multiple
of $16$ are indeed sparse, in accord with the second observation
above.

As to the first result, it exhibits an important dichotomy for RCFTs:
a model has either one non-trivial Galois current, or it has none.
As we shall see in the next section, the existence of a non-trivial
Galois current has important consequences regarding the structure
of the modular representation, so the above mentioned dichotomy is
not only meaningful, but also relevant. In this respect one might
say that the existence of a non-trivial Galois current for the Ising
model is a good signal of its special nature.

\section{Ordinary vs projective kernel}

Given a linear representation of a group, it is natural to look at
the kernel of the representation, i.e. the set of group elements that
are represented by the identity operator. As we have mentioned in
section 2, in case of the modular representation associated to a RCFT,
the kernel is a congruence subgroup whose level equals the order $\FA{}$
of $T$. Knowing this fundamental result, the next natural step is
to look for a simple description of the kernel. As it turns out the
natural object to look at is not the kernel itself, but rather the
\emph{projective kernel} $\FB{}=\set{m\in\Gamma\left(1\right)}{D\left(m\right)=\xi\left(m\right)1}$,
which consists of those modular transformations that are represented
by scalar matrices, where $\xi:\FB{}\rightarrow\mathbb{C}$ is a linear
character of $\FB{}$, which we term the \emph{central character}..
Let's explain why this is so.

Consider a composite system denoted $\mathcal{C}_{1}\otimes\mathcal{C}_{2}$,
which is made up of two independent and non-interacting subsystems
$\mathcal{C}_{1}$ and $\mathcal{C}_{2}$. Note that if both subsystems
$\mathcal{\mathcal{C}}_{1}$and $\mathcal{C}_{2}$ are RCFTs, then
so is their composite $\mathcal{C}_{1}\otimes\mathcal{C}_{2}$. As
the subsystems are completely independent, one expects that it should
be possible to determine unambiguously the value of any ''natural''
quantity for the composite system from the knowledge of the corresponding
values for the subsystems.

In this respect, the partition function and all correlators are ''natural'',
for they factorize in the composite system into a product of corresponding
quantities for the subsystems. Similarly, the modular representation
is ''natural'', for the modular representation associated to $\mathcal{C}_{1}\otimes\mathcal{C}_{2}$
is the tensor product of the modular representations associated to
$\mathcal{C}_{1}$ and $\mathcal{C}_{2}$. On the other hand, the
kernel is not ''natural'', for one cannot determine the kernel of
a tensor product from the sole knowledge of the kernels of the factors.
The ''natural'' notion is that of the projective kernel, because
the projective kernel of a tensor product is simply the intersection
of the projective kernels of the factors. This is the conceptual explanation
for looking at the projective kernel instead of the ordinary one.
We note that the same ''naturality'' argument explains why it is
the parameter $K$ and not the conductor $N$ that enters most of
our results, for the former is ''natural'' in the above sense, while
the later is not. 

After these preliminaries, let's describe the structure of the projective
kernel. To this end, let's first introduce the notation \[
\FD{g}=\left\{ \left(\begin{array}{cc}
a & b\\
c & d\end{array}\right)\in\Gamma\left(1\right)\,|\, a,d\in\mathfrak{g},\,\,\Mod{b,c}{0}{K}\right\} \]
for a subgroup $\mathfrak{g}<\FC{K}$. Clearly, $\Gamma\left(K,\left\{ 1\right\} \right)$
equals the principal congruence subgroup $\Gamma\left(K\right)$,
from which follows that $\FD{g}$ is a congruence subgroup for any
$\mathfrak{g}<\FC{K}$. 

Remember the Galois permutations $\pi_{l}$ from section 2, and consider
the kernel of the permutation action $\pi$, i.e. the set $\FE{}{}=\set{l\in\FC{\FA{}}}{\pi_{l}=1}$.
Clearly, $\FE{}{}$ is a subgroup of $\FC{\FA{}}$, and reducing all
elements of $\FE{}{}$ modulo $K$ we get a subgroup $\mathfrak{h}$
of $\FC{K}$. It does follow from general properties of the Galois
action that the exponent of $\mathfrak{h}$ is $2$, in other words
$\Mod{l^{2}}{1}{K}$ for all $l\in\mathfrak{h}$ \cite{CMP}. The
basic result about the projective kernel is that $\FD{h}$ is a normal
subgroup of $\FB{}$ - in particular $\FB{}$ is a congruence subgroup
of level $K$ -, and there is an explicit isomorphism between the
factor group \[
\FB{}/\FD{h}\]
and the group of Galois currents! This means that the knowledge of
$K$, the subgroup $\mathfrak{h}$ and the group of Galois currents
does completely determine the projective kernel. Note that the index
of the principal congruence subgroup $\Gamma\left(K\right)$ in $\FB{}$,
which equals the number of Galois currents times the order of $\mathfrak{h}$,
is always a power of $2$ by the above, a result that seems fairly
non-trivial.

As an example, let's consider once again the Ising model $\mathcal{M}\left(4,3\right)$.
As discussed in section 3, one has $e=3$ and $K=16$, and there is
a non-trivial Galois current (the energy operator $\varepsilon$).
In this case, the group $\mathfrak{h}$ consists of the prime residues
$\left\{ 1,7,9,15\right\} $ modulo $16$. As to the projective kernel,
it has the coset decomposition \[
\FB{}=\FD{h}\cup\left(\begin{array}{cc}
3 & 8\\
8 & 11\end{array}\right)\FD{h}\]

\section{Discussion}

As we have seen, the notion of Galois currents is not only meaningful
in the sense that there exists non-trivial examples, but also relevant
to the analysis of the properties of RCFTs. While we have concentrated
on their impact on the structure of the projective kernel, it is quite
plausible that they play an important role for other aspects of the
theory as well, e.g. fusion rules and modular invariants. Elucidating
these connections seems to be a rewarding task for the future. 

The simple characterization of the projective kernel described in
section 4 should be regarded only as a first step. It should be followed
by an understanding of the central character, the map $\xi:\FB{}\rightarrow\mathbb{C}$
appearing in the definition of the projective kernel. Another interesting
point would be to find out which subgroups $\mathfrak{h}<\FC{K}$
are allowed for a given value of $K$.

Finally, one might speculate on the relevance of the above for the
classification of RCFTs. The knowledge of the projective kernel might
be an important ingredient of classification attempts. From another
point of view, one may hope that the special properties of RCFTs with
non-trivial Galois currents could lead to a classification of this
special class of model.


\begin{thebibliography}{10}
\bibitem{Cardy}J. Cardy, Nucl. Phys. \textbf{B270}, 186 (1986). 
\bibitem{Ver}E. Verlinde, Nucl. Phys. \textbf{B300}, 360 (1988). 
\bibitem{MS}G. Moore and N. Seiberg, Commun. Math. Phys. \textbf{123}, 177 (1989). 
\bibitem{Turaev}V.G. Turaev, \emph{Quantum Invariants of Knots and 3-Manifolds} (de
Gruyter, New York 1994). 
\bibitem{MTC}B. Bakalov and A.A. Kirillov, \emph{Lectures on Tensor Categories
and Modular Functors} (AMS, Providence 2000). 
\bibitem{FRS}J. Fuchs, I. Runkel, C. Schweigert, hep-th/0204148 .
\bibitem{Moore}G. Moore, Nucl. Phys. \textbf{B293}, 139 (1987). 
\bibitem{Eholzer}W. Eholzer, Commun. Math. Phys. \textbf{172}, 623 (1995), hep-th/9408160
. 
\bibitem{Eh-Sko}W. Eholzer and N.-P. Skoruppa, Commun. Math. Phys. \textbf{174}, 117
(1995), hep-th/9407074 . 
\bibitem{BCIR}M. Bauer, A. Coste, C. Itzykson and P. Ruelle, J. Geom. Phys. \textbf{22},
134 (1997), hep-th/9604104 .
\bibitem{CG2}A. Coste and T. Gannon, math-QA/9909080. 
\bibitem{CMP}P. Bantay, Commun. Math. Phys. \textbf{233}:423-438 (2003), math-QA/0102149
.
\bibitem{sc1}A.N. Schellekens, S. Yankielowicz, Phys.Lett.\textbf{B227}:387 (1989).
\bibitem{sc2}K.A. Intriligator, Nucl.Phys.\textbf{B332}:541 (1990 ).
\bibitem{sc3}A.N. Schellekens, S. Yankielowicz, Int.J.Mod.Phys.\textbf{A5}:2903-2952
(1990).
\bibitem{BG}J. de Boere and J. Goeree, Commun. Math. Phys. \textbf{139}, 267 (1991). 
\bibitem{CG1}A. Coste and T. Gannon, Phys. Lett. \textbf{B323}, 316 (1994). \end{thebibliography}
\end{document}